\documentclass[journal,twocolumn,10pt]{IEEEtran}
\usepackage{textcomp}
\usepackage{stfloats}
\usepackage{url}
\usepackage{graphicx}
\usepackage{verbatim}
\usepackage{upgreek}
\usepackage{amssymb,amsmath}
\usepackage{color}
\usepackage{epstopdf}
\usepackage{bm}
\usepackage{cite}
\usepackage{array,color}
\usepackage{algorithm}
\usepackage{algpseudocode}
\usepackage{amsmath}
\usepackage{amsfonts}
\usepackage{amsthm}
\usepackage{graphics}
\usepackage{epsfig}
\usepackage{soul}
\usepackage{gensymb}
\soulregister\cite7
\soulregister\ref7
\usepackage{subfigure}

\usepackage{setspace}

\usepackage[colorlinks=true, linkcolor=blue, citecolor=blue, urlcolor=blue]{hyperref}

\def\BibTeX{{\rm B\kern-.05em{\sc i\kern-.025em b}\kern-.08em
		T\kern-.1667em\lower.7ex\hbox{E}\kern-.125emX}}

\begin{document}

\title{Toward Low-Altitude Embodied Intelligence: A Sensing-Communication-Computation-Control Closed-Loop Perspective}

\author{Jihao Luo, Zesong Fei,~\IEEEmembership{Senior~Member,~IEEE}, Xinyi Wang,~\IEEEmembership{Member,~IEEE}, Shuntian Tang, 
	
	Zilong Liu,~\IEEEmembership{Senior~Member,~IEEE}, Yiqing Zhou,~\IEEEmembership{Senior~Member,~IEEE}
	

	\thanks{Jihao Luo, Zesong Fei, Xinyi Wang and  Shuntian Tang are with the School of Information and Electronics, Beijing Institute of Technology, Beijing 100081, China (e-mail: jihaoluo$\_$bit@icloud.com, feizesong@bit.edu.cn, wangxinyi@bit.edu.cn, 3120255634@bit.edu.cn).}
	\thanks{Zilong Liu is with the School of Computer Science and Electronic Engineering, University of Essex, UK (email: zilong.liu@essex.ac.uk).}
	\thanks{Yiqing Zhou is with the University of Chinese Academy of Sciences; State Key Laboratory of Processors, Institute of Computing Technology, Chinese Academy of Sciences, Beijing 100190, China	(e-mail: zhouyiqing@ict.ac.cn).}
	

}



\maketitle

\begin{abstract}
	The rapid growth of the low-altitude economy drives increasingly autonomous unmanned aerial vehicle (UAV) operations, giving rise to low-altitude embodied intelligence (LAEI), in which sensing, communication, computation, and control (SC$^3$) are tightly integrated to enable closed-loop interaction, ensuring timely, effective, and safe responses in complex or unknown environments. This article systematically explores the LAEI networks, from its fundamental architecture to the diverse scenarios that it can support. We examine key enabling techniques that sustain timely information exchange and effective decision feedback within the $\text{SC}^3$ closed loop. A representative low-altitude UAV mission in an unknown urban area is presented as a case study, where the UAV provides communication services and performs environmental sensing to inform closed-loop control, illustrating how coordinated $\text{SC}^3$ capabilities enable efficient and responsive operation. By identifying major challenges and outlining future research directions, this work serves as a cornerstone for developing next-generation low-altitude intelligent systems.
\end{abstract}


\section{Introduction}
The rapid emergence of low-altitude economy is driving an unprecedented demand for intelligent, autonomous, and reliable aerial services \cite{10049809}. Unmanned aerial vehicles (UAVs) have become key operational platforms in low-altitude airspace, supporting a broad range of applications, such as environmental monitoring, emergency response, and urban inspection \cite{7470933}. By operating in close proximity to the ground, UAVs are able to perceive fine-grained environmental information and deliver flexible, on-demand services that might be difficult to achieve with conventional terrestrial or high-altitude platforms. 

Despite these advantages, low-altitude environment poses significant challenges to UAV autonomy and system reliability. Specifically, dense obstacles, heterogeneous user demands, and rapidly varying air-ground channels jointly characterize low-altitude airspace \cite{luo2025trajectory}. To operate safely and efficiently in these environments, UAVs must maintain continuous situational awareness, interpret changing conditions, and adapt their behavior in real time, making embodied intelligence a core capability for reliable and scalable low-altitude operations.

In general, embodied intelligence follows a closed-loop paradigm of sensing, computation, and control, in which onboard sensors collect environmental data, onboard processors perform inference and decision-making, and control modules execute corresponding actions \cite{10926861}. This tightly coupled perception-action loop implicitly assumes that all intelligence can be realized locally. However, in low-altitude scenarios, the complex environment and high dimensionality of sensory inputs significantly increase the computational demands. Constrained by the payload, energy, and onboard processing capabilities, purely onboard implementations often struggle to support timely inference and adaptive control, thus limiting the autonomy performance in practical deployments.

To address these limitations, this article examines a sensing-communication-computation-control ($\text{SC}^3$) closed-loop perspective for enabling embodied intelligence in low-altitude UAV networks, whereby low-altitude wireless networks (LAWNs) are conceived to integrate air-ground communication, computation, and coordination for flexible, autonomous UAV operations. In particular, communication is treated as an inherent component of the perception-action loop, enabling UAVs to interact with distributed ground-based computing resources. The sensed environmental information can be conveyed through air-ground links for real-time inference and policy adaptation, while updated control decisions are returned to UAVs to guide their actions. By integrating sensing, communication, computation, and control into a unified closed-loop architecture, the $\text{SC}^3$ framework provides a holistic perspective on adaptive and autonomous UAV operations under stringent latency and reliability constraints in complex low-altitude environments, while offering practical guidance for the design and deployment of future LAEI systems.

Building on this perspective, we provide an overview for the low-altitude embodied intelligence (LAEI) networks, ranging from its system architecture, representative applications, critical techniques, to remaining challenges. The main contributions are summarized as follows:
\begin{itemize}
	\item We propose an $\text{SC}^3$ closed-loop perspective for LAEI, and present a hierarchical UAV--edge--cloud architecture with representative low-altitude scenarios to illustrate the operating mechanism of LAEI networks.
	\item We summarize the critical techniques for implementing the $\text{SC}^3$ loop, including mobility-enabled high-resolution sensing for environmental awareness, Doppler-resilient air--ground communications, hierarchical computation orchestration across the UAV--edge--cloud continuum, and reinforcement learning (RL)-enabled adaptive control.
	\item We provide a representative case study of a low-altitude UAV mission in an unknown urban area to demonstrate how the integrated $\text{SC}^3$ capabilities translate into improved mission completion efficiency, and finally outline open challenges and future research directions toward scalable and resilient LAEI networks.
\end{itemize}

\section{System Architecture}

\begin{figure*}[htbp]
	\centering
	\includegraphics[width=0.7\textwidth,height=0.4\textheight, keepaspectratio]{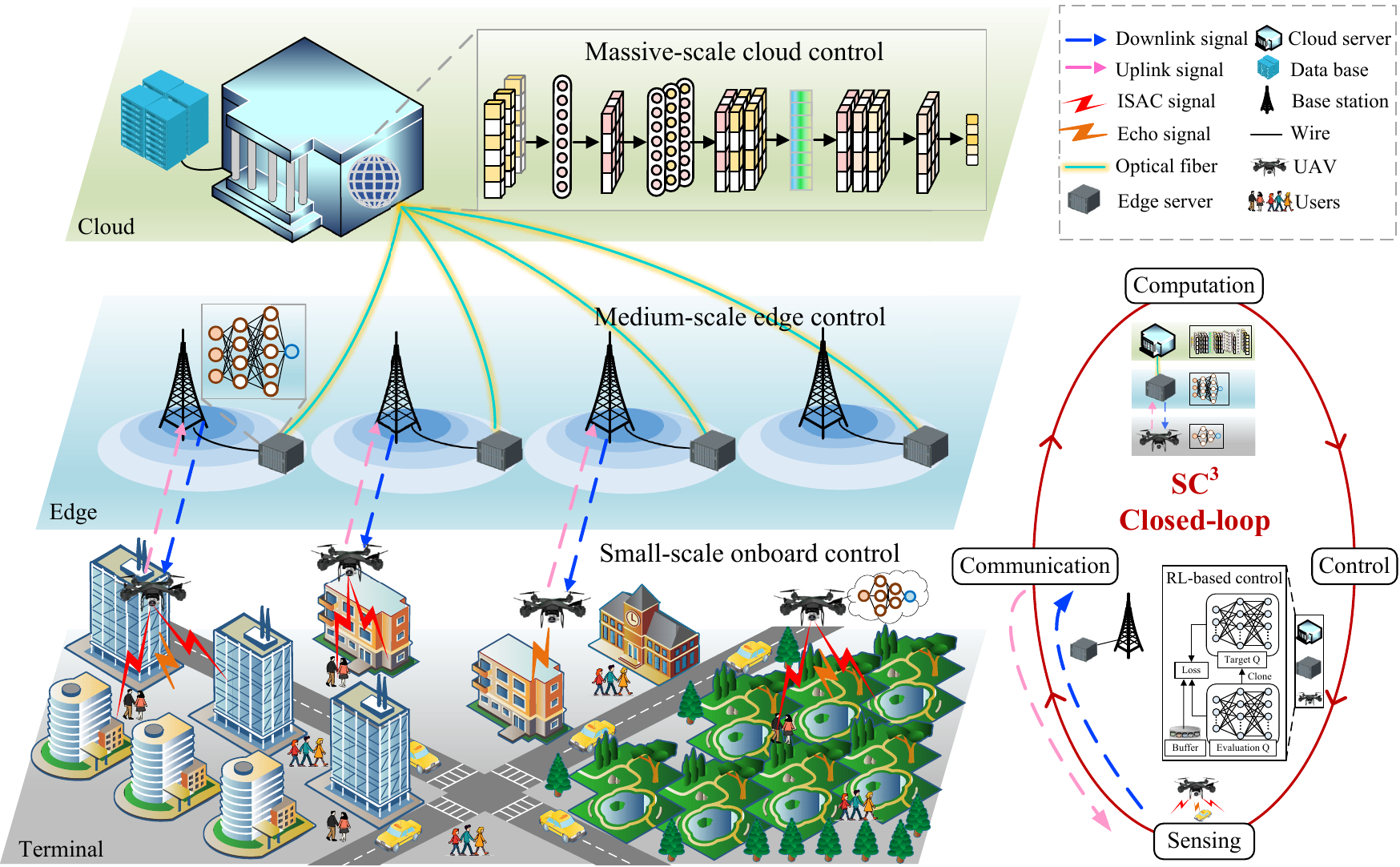}
	\caption{System architecture and operating mechanism for $\text{SC}^3$ based LAEI networks.}
	\label{system_model}
	\vspace{-0.3cm}
\end{figure*}

Fig.~\ref{system_model} illustrates the architectural framework of the LAEI network, where specific roles are assigned to UAVs, edge servers, and cloud platforms to establish a hierarchical $\text{SC}^3$ loop. UAVs serve as the embodied agents that directly interact with the physical environment, while edge and cloud nodes provide tiered support for scalable computing and coordination.

During flight, each UAV transmits integrated sensing and communication (ISAC) signals, simultaneously supporting wireless connectivity and environmental perception. With the lightweight onboard models, UAVs process echo signals locally for mission-critical and time-sensitive tasks such as urgent obstacle avoidance. On the other hand, raw or pre-processed data requiring granular analysis, such as detailed environmental mapping, building distribution analysis, and target dynamics tracking, are transmitted through network to edge and cloud nodes.

Edge servers function as intermediate processing hubs, handling data offloaded from UAVs to facilitate timely control and decision assistance. By relieving UAVs of heavy computational burdens, edge servers preserve the aerial agents' mobility and real-time responsiveness, effectively serving as the primary engines for latency-sensitive tasks. Furthermore, through high-capacity backhaul connections, edge servers interface seamlessly with cloud platforms, ensuring that local decisions are augmented by broader system intelligence.

Complementing the edge, the cloud platform provides system-level orchestration. By aggregating information from multiple UAVs and edge servers, the cloud executes computationally intensive and latency-tolerant tasks, such as long-horizon planning, large-scale optimization, and cross-region coordination. Operating on a macro-temporal and spatial scale, the cloud generates high-level directives that are disseminated through edge servers to UAVs, aligning local actions with overarching system objectives.

Ubiquitous communication links sustain the bidirectional flow of sensing information and control commands across this multi-tier architecture, effectively binding sensing, communication, computation, and control into a closed loop. This hierarchical integration enables UAVs to respond adaptively to real-time observations, accumulated knowledge, and distributed guidance, thereby facilitating robust embodied intelligence in complex low-altitude environments. Under this $\text{SC}^3$ framework, different application scenarios impose distinct requirements on reliability, latency, throughput, etc., giving rise to heterogeneous operational needs. To capture these differences, Table~\ref{use-cases} summarizes four representative scenarios selected to reflect diverse low-altitude environments and service requirements. These examples illustrate how LAEI networks address varied performance demands while enabling environment-aware operations in dynamic low-altitude environments.

\begin{table*}[htbp]
	\centering
	\caption{Potential Use Cases of LAEI Network}
	\setstretch{1.2}
	\label{use-cases}
	\resizebox{\textwidth}{!}{
		\begin{tabular}{|l|l|l|}
			\hline
			\multicolumn{1}{|c|}{\textbf{Use Cases}} &
			\multicolumn{1}{c|}{\textbf{Key Requirements and Features}} &
			\multicolumn{1}{c|}{\textbf{Merits of Deploying LAEI Networks}} \\
			\hline
			
			\textbf{Urban hotspots} &
			\begin{tabular}[c]{@{}l@{}}
				$\bullet$ Throughput and spectral efficiency are primary, with fluctuating service loads \\
				$\bullet$ High-density users with rapidly changing traffic demands \\
				$\bullet$ Complex low-altitude airspace with rich obstacles and strong co-channel interference
			\end{tabular} &
			\begin{tabular}[c]{@{}l@{}}
				$\bullet$ Provide flexible and high-capacity communication support \\ 
				$\bullet$ Adapt swiftly to dynamic user distribution and service load \\ 
				$\bullet$  Enhance operational reliability through real-time low-altitude sensing \end{tabular}\\
			\hline
			
			\textbf{Disaster areas} &
			\begin{tabular}[c]{@{}l@{}}
				$\bullet$ Reliability and latency are critical for emergency alerts and coordination \\
				$\bullet$ Ground infrastructure damaged or unavailable, with highly uncertain environments \\
				$\bullet$ Dynamic rescue needs requiring rapid situational updates and robust links
			\end{tabular} &
			\begin{tabular}[c]{@{}l@{}} 
				$\bullet$ Enable rapid restoration of emergency communication coverage \\ 
				$\bullet$ Improve situational awareness via low-altitude sensing \\ 
				$\bullet$ Support coordinated rescue operations through reliable connectivity
			\end{tabular} \\
			\hline
			
			\textbf{Remote regions} &
			\begin{tabular}[c]{@{}l@{}}
				$\bullet$ Coverage and energy efficiency dominate, while throughput demands are moderate \\
				$\bullet$ Sparse population with low-density and uneven service needs \\
				$\bullet$ Limited or nonexistent infrastructure with challenging terrain and coverage gaps
			\end{tabular} &
			\begin{tabular}[c]{@{}l@{}} 
				$\bullet$ Offer cost-effective and easily deployable connectivity \\ 
				$\bullet$ Improve coverage reach through flexible low-altitude operations \\ 
				$\bullet$ Support long-term monitoring and communication in isolated areas \end{tabular} \\
			\hline
			
			\textbf{Conflict zones} &
			\begin{tabular}[c]{@{}l@{}}
				$\bullet$ Ultra-high reliability and ultra-low latency are simultaneously required \\
				$\bullet$ Time-critical UAV missions with high operational risks and limited prior awareness \\
				$\bullet$ Communication links vulnerable to jamming, obstruction, or disruption
			\end{tabular} &
			\begin{tabular}[c]{@{}l@{}} 
				$\bullet$ Provide resilient and mobile communication support \\ 
				$\bullet$ Strengthen environmental awareness through continuous sensing \\ 
				$\bullet$ Enable fast coordination and decision support for ongoing missions \end{tabular} \\
			\hline
			
	\end{tabular}}
\end{table*}

\section{Critical Techniques in LAEI Network}

Effective LAEI hinges on the seamless integration of sensing, communication, computation, and control. In this section, we delineate the critical technologies within each domain that facilitate the $\text{SC}^3$ closed loop.

\subsection{Joint Synthetic Aperture Radar and Communications}
\begin{figure}
	\centering
	\includegraphics[width=3.6 in]{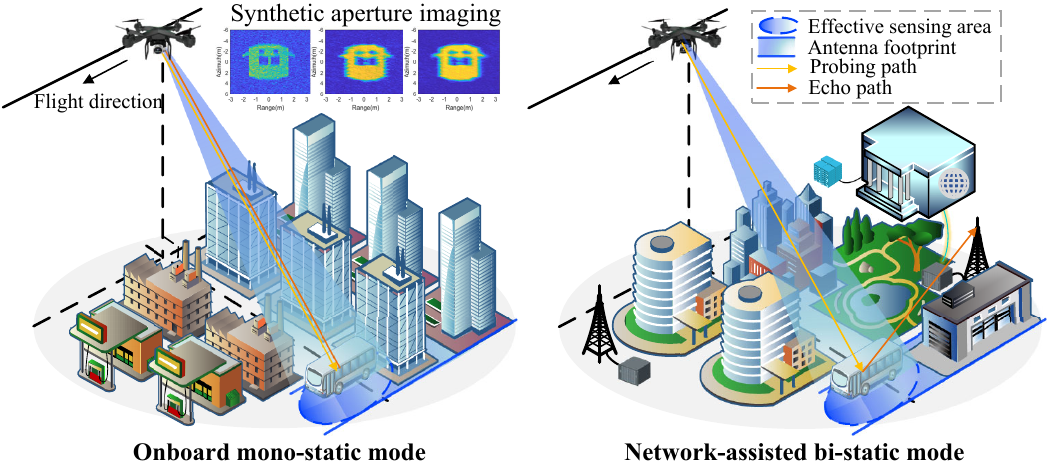}
	\caption{Illustrations of two JSARC modes in the LAEI networks.}
	\label{sensingtechnique}
	\vspace{-0.3cm}
\end{figure}

While providing ubiquitous communication coverage for ground users (GUs), UAVs are required to perceive their surroundings for environment learning and trajectory design. Due to strict payload and power constraints, UAVs often lack the large antenna arrays required for high-resolution imaging. Joint synthetic aperture radar and communications (JSARC) elegantly bypasses this hardware limitation by exploiting UAVs' flexible mobility. Specifically, by collecting echoes of the transmitted communication signals along its trajectory, the UAV is able to synthesize a large virtual aperture to achieve fine-grained resolution. As illustrated in Fig. \ref{sensingtechnique}, JSARC can be categorized into two primary paradigms: onboard mono-static JSARC and network-assisted bi-static JSARC modes.

\textit{1) Onboard Mono-Static JSARC}

Mono-static JSARC treats the UAV as a self-contained entity that simultaneously transmits communication signals and processes the echoes. In this mode, the UAV reuses the communication waveform, e.g., orthogonal frequency division multiplexing (OFDM), as a radar probe. As the UAV traverses its path, echoes collected at successive intervals are coherently combined to synthesize a virtual aperture significantly larger than the physical onboard antenna. This tight coupling between mobility and sensing allows for rapid, autonomous situational awareness, making it well suited for time-critical tasks such as local obstacle detection and real-time scene mapping.

To maintain imaging quality under continuous motion, mobility-aware signal processing is required to compensate for trajectory-induced phase variations, ensuring that the echoes are accurately mapped into range-azimuth domains \cite{10715676}. Geometry-aware operations, such as trajectory-aligned resampling, rotation recovery, and coherent echo accumulation, can further enhance sensing quality \cite{11244068}. While onboard SAR preserves sensing autonomy, it imposes a high computational burden on the UAV, necessitating a balance between sensing resolution and onboard energy consumption.

\textit{2) Network-Assisted Bi-Static JSARC}

To alleviate the requirements of onboard computation, network-assisted bi-static JSARC decouples signal transmission and echo reception across the air-ground interface. In this mode, the UAV serves as a mobile transmitter providing communication coverage, while ground-based stations, e.g., BSs capture the reflected echoes. By offloading echo reception and heavy SAR signal processing to the network side, this mode significantly reduces the UAV's ``Size, Weight, and Power" (SWAP) burden, while preserving the high-resolution benefits of the synthetic aperture.

Network-assisted JSARC allows for more sophisticated imaging algorithms that would be prohibitive onboard. For example, ground-based back-projection processing can effectively handle complex air–ground geometries and compensate for non-linear flight trajectories \cite{11271028}. The primary challenge in this networked mode is high-precision synchronization: Achieving phase alignment between the mobile UAV transmitter and receiver is essential for coherent SAR imaging. Advanced synchronization schemes based on direct-path reference signals lay the foundation of stable echo accumulation over extended apertures. Ultimately, networked JSARC enables large-scale, high-fidelity environmental perception by fusing information from multiple perspectives across the LAWNs.

\subsection{Reliable Communication over High-Mobility Air-Ground Links}\label{teccomm}
In the LAEI framework, communication acts as the nervous system, bridging mobile agents with distributed computation and control hubs. Compared with conventional aerial communications, the LAEI closed loop demands a dual-functional communication design to accommodate two distinct flows: A high-capacity uplink for offloading voluminous raw sensing data and a high-reliability and low-complexity downlink for delivering time-sensitive control commands. Note that both of them must operate under the severe Doppler shifts and fast-fading characteristics of air-ground channels.

\textit{1) Sensing Data Uploading: High-Capacity $\text{\&}$ Doppler Resilience}

The primary objective of the air-to-ground uplink is to transfer rich sensing data to edge servers for intensive processing. Under high-speed flight, traditional OFDM suffers from severe inter-carrier interference (ICI), which significantly degrades transmission throughput. To combat this challenge, affine frequency division multiplexing (AFDM) \cite{10087310} has emerged as a solution. By modulating data in the affine frequency (AF)-domain, AFDM effectively converts the doubly-dispersive channel into a quasi-static representation, thereby enhancing Doppler resilience. A major advantage of AFDM is that it enjoys full channel diversity by spreading every data symbol in the time-frequency domain. At the same time, one can easily configure AFDM system parameters to include OFDM as a special case.  

However, AFDM typically requires complex receiver processing and introduces tighter constraints on resource mapping and scheduling. A more pragmatic architecture for LAEI involves cross-domain signal processing: Utilizing AF-domain pilots to acquire accurate channel state information (CSI) while performing low-complexity equalization in the time-frequency (TF)-domain. Additionally, the resulting AF-domain channel matrices are proven to exhibit autoregressive behavior across adjacent symbols \cite{shuntian2026doppler}, which can be leveraged by learning-based predictors, such as long short-term memory networks (LSTM), to anticipate channel evolution, ensuring robust uplink connectivity even during aggressive maneuvers. Table \ref{ofdm-afdm-comparison} provides a comparative analysis of OFDM and AFDM across several technical dimensions, highlighting the performance-complexity synergy achieved by integrating AF-domain pilots into the OFDM framework.

\textit{2) Control Signaling Delivering: Reliability with Receiver Simplicity}

Ground-to-air downlink communication delivers decision and control commands from edge servers to UAV platforms; hence, the priority shifts from high throughput to ultra-reliable and low-latency command dissemination. Unlike uplink transmission that can tolerate buffering and retransmission, delayed or misinterpreted downlink commands can lead to catastrophic failures in autonomous navigation. Consequently, the UAV receiver has to recover control information rapidly and reliably under fast-varying air-ground channels.

A fundamental challenge in this context is the stark computational disparity between the network nodes: while the ground-based edge infrastructure possesses abundant computing capability, the UAV is governed by stringent SWAP constraints. To bridge this gap, a practical design philosophy is therefore to transfer the signal processing burden from the UAV receiver to the edge transmitter. By exploiting channel reciprocity to perform sophisticated pre-equalization at the edge transmitter, symbol-level precoding enables the UAV to recover control signaling via direct symbol detection, entirely bypassing the energy-intensive channel estimation and equalization at the UAV receiver. Furthermore, recent progress in sparse Bayesian learning-based AFDM channel estimation \cite{Tang2025} enables accurate CSI acquisition at the edge based on the received AF-domain pilots, ensuring symbol-level precoding remains robust even under high mobility.

\begin{table*}
	\centering
	\caption{Comparison of OFDM, AFDM, and OFDM with AF-Domain Pilot over Air-Ground Communications}
	\setstretch{1.2}
	\label{ofdm-afdm-comparison}
	\resizebox{\textwidth}{!}{
		\begin{tabular}{|l|l|l|l|}
			\hline
			\multicolumn{1}{|c|}{\textbf{Characteristics}} & 
			\multicolumn{1}{c|}{\textbf{OFDM}} & 
			\multicolumn{1}{c|}{\textbf{AFDM}} &
			\multicolumn{1}{c|}{\textbf{OFDM with AF-Domain Pilot}} \\ \hline
			
			\textbf{Transformation Domain} & 
			\begin{tabular}[c]{@{}l@{}}
				$\bullet$ TF-domain 
			\end{tabular} & 
			\begin{tabular}[c]{@{}l@{}}
				$\bullet$ Time-AF domain 
			\end{tabular} &
			\begin{tabular}[c]{@{}l@{}}
				$\bullet$ TF-domain data transmission with AF-domain pilot processing 
			\end{tabular} \\ \hline
			
			\textbf{Channel Estimation} & 
			\begin{tabular}[c]{@{}l@{}}
				$\bullet$ Conventional pilot-based channel estimation \\
				$\bullet$ Degradation in fast-varying channels
			\end{tabular} & 
			\begin{tabular}[c]{@{}l@{}}
				$\bullet$ AF-domain channel estimation \\
				$\bullet$ Effective CSI acquisition in doubly selective channels
			\end{tabular} &
			\begin{tabular}[c]{@{}l@{}}
				$\bullet$ AF-domain channel estimation and transformation to TF-domain\\
				$\bullet$ Improved CSI accuracy under high mobility
			\end{tabular} \\ \hline
			
			\textbf{Equalization Complexity} & 
			\begin{tabular}[c]{@{}l@{}}
				$\bullet$ One-dimensional per-subcarrier equalization \\
				$\bullet$ Low computational complexity
			\end{tabular} & 
			\begin{tabular}[c]{@{}l@{}}
				$\bullet$ Two-dimensional equalization in AF-domain \\
				$\bullet$ Higher computational complexity
			\end{tabular} &
			\begin{tabular}[c]{@{}l@{}}
				$\bullet$ One-dimensional equalization aided by AF-domain channel estimation \\
				$\bullet$ Moderate computational complexity
			\end{tabular} \\ \hline
			
			\textbf{Doppler Impact} & 
			\begin{tabular}[c]{@{}l@{}}
				$\bullet$ Sensitive to Doppler-induced ICI \\
				$\bullet$ Performance degradation under high mobility
			\end{tabular} & 
			\begin{tabular}[c]{@{}l@{}}
				$\bullet$ Doppler-induced coupling naturally accommodated\\
				$\bullet$ Robust performance against high mobility
			\end{tabular} &
			\begin{tabular}[c]{@{}l@{}}
				$\bullet$ Improved Doppler awareness via AF-domain pilots \\
				$\bullet$ Enhanced robustness compared with conventional OFDM
			\end{tabular} \\ \hline
			
			\textbf{Resource Mapping} & 
			\begin{tabular}[c]{@{}l@{}}
				$\bullet$ Flexible TF-domain mapping \\
				$\bullet$ Supports fine-grained scheduling
			\end{tabular} & 
			\begin{tabular}[c]{@{}l@{}}
				$\bullet$ Full-band coupling despite full diversity gain \\
				$\bullet$ Less flexible for fine-grained scheduling
			\end{tabular} & 
			\begin{tabular}[c]{@{}l@{}}
				$\bullet$ Preserves TF-domain mapping flexibility \\
				$\bullet$ Supports fine-grained scheduling
			\end{tabular} \\ \hline
			
		\end{tabular}
	}
	\vspace{-0.3cm}
\end{table*}

\subsection{Hierarchical Computation Orchestration}

\begin{figure}[htbp]
	\centering
	\includegraphics[width=3.6 in]{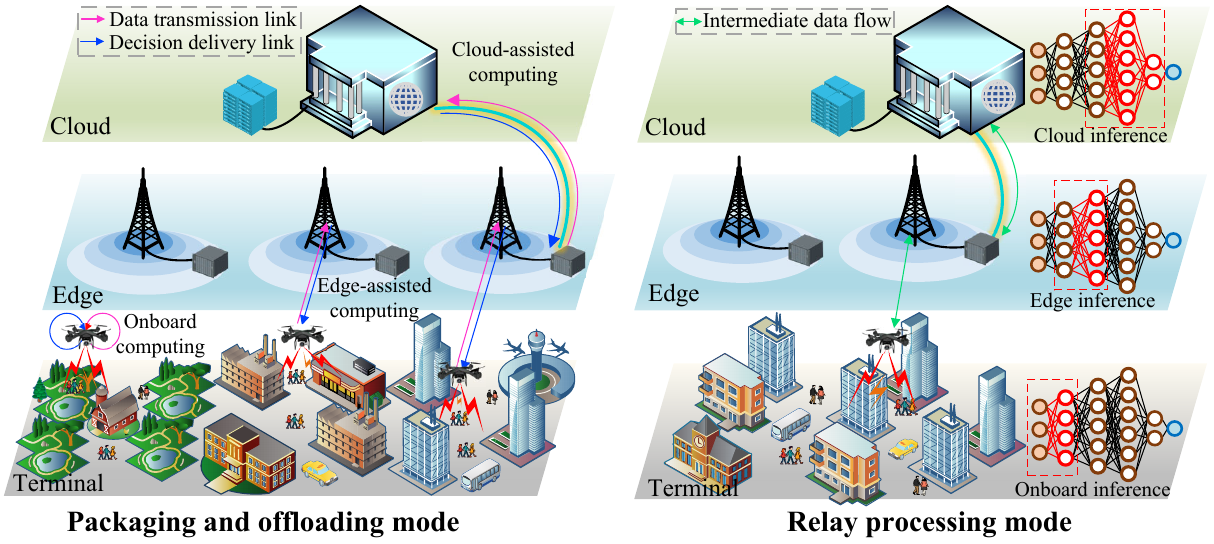}
	\caption{Illustrations of hierarchical computation orchestration in LAEI networks.}
	\label{computationtechnique}
	\vspace{-0.3cm}
\end{figure}

To accommodate the diverse latency and complexity requirements of perception-centric missions, the LAEI framework moves beyond static execution patterns toward a hierarchical computation orchestration paradigm. This architecture adaptively balances latency, energy, and accuracy by dynamically mapping computational workloads across the UAV-edge-cloud continuum. As illustrated in Fig. \ref{computationtechnique}, the orchestration strategies can be categorized into two paradigms: holistic task offloading and cross-layer split computing.

\textit{1) Holistic Task Offloading: Packaging and Offloading Mode}

The holistic offloading paradigm treats the computational task as a whole, assigning its execution to either local UAV, an edge server, or a central cloud, according to the available resources. Lightweight tasks, such as reactive obstacle avoidance, are handled locally to minimize latency; whereas resource-intensive workloads, such as large-scale environmental mapping, are offloaded to more powerful external platforms. 

Crucially, in large-scale LAEI deployments, offloading decisions are not solely dictated by a single task's complexity or latency requirements. Instead, heterogeneous characteristic of diverse missions, ranging from jitter-sensitive control to high-resolution imaging, must be accounted for while performing network-wide resource orchestration. By taking a holistic view on the aggregate workload across the UAV-edge-cloud continuum, the system can prevent localized congestion at edge nodes and ensure the resources to be allocated to the most critical tasks. As illustrated in \cite{10966042}, employing alternating optimization for multi-dimension offloading design can significantly reduce execution latency and energy consumption, preserving the UAVs' limited battery life for essential flight operations while maintaining robust system-wide responsiveness.

\textit{2) Cross-Layer Split Computing: Relay Processing Mode}

Unlike the binary decision-making of holistic offloading, split computing exploits the inherent modularity of deep neural networks to distribute different stages of a computation task across the network. By partitioning a perception model at specific ``bottleneck" layers, computation is executed in a staged pipeline, i.e., the UAV performs initial feature extraction, the edge server handles intermediate inference, and the cloud completes final refinement. The key advantage of this ``relay processing" mechanism is the transmission of compact intermediate features instead of raw sensing data, which significantly alleviates the pressure on constrained air-ground fronthaul. A representative realization on such relay processing based integrated sensing, communication, and computing system has been investigated in \cite{11226946}. By partitioning computation across bottleneck layers and relaying compact intermediate features, the relay processing mode reduces fronthaul load and end-to-end latency, making it well suited for perception-centric LAEI applications with stringent real-time requirements.

\subsection{Reinforcement Learning-Enabled Adaptive Control}

The ultimate objective of the LAEI framework is to close the loop through adaptive control, enabling UAV to translate environmental awareness into precise, goal-oriented maneuvers. Achieving this demands continuous assimilation of real-time sensing observations and network feedbackwithin a closed-loop orchestration. However, the intrinsic complexity of this paradigm—characterized by the high dimensionality of state and action spaces as well as the tight coupling of sensing, communication, and motion dynamics—renders conventional model-based or rule-based control strategies unstable and insufficient.

RL emerges as a compelling model-free paradigm by casting UAV control as a sequential decision-making process driven by direct interaction with the environment. Through navigating the intricate state-action spaces via iterative trial-and-error exploration, RL agents refine robust control policies that balance multiple objectives—such as flight safety, sensing accuracy, and communication reliability—without relying on explicit analytical models. This adaptability makes RL particularly suited for LAEI scenarios, where environmental dynamics are unpredictable. For instance, in trajectory planning, RL enables UAVs to learn optimal flight paths directly from real-time feedback, incorporating factors like obstacle distributions and time-varying air-ground channels  \cite{11149303}. Beyond path planning, deep RL has been employed in edge-assisted multi-UAV networks to jointly optimize task offloading and resource allocation—including CPU frequencies, bandwidth, and transmission power—while efficiently handling hybrid discrete-continuous action spaces to minimize long-term delay and energy consumption \cite{10473128}. Consequently, by fusing multi-source observations into responsive control actions, RL serves as the ``intelligence engine" that empowers UAVs with autonomous decision-making capabilities, fulfilling the vision of embodied intelligence within the unified $\text{SC}^3$ framework.

\begin{figure*}[htbp]
	\centering
	\includegraphics[width=0.95\textwidth]{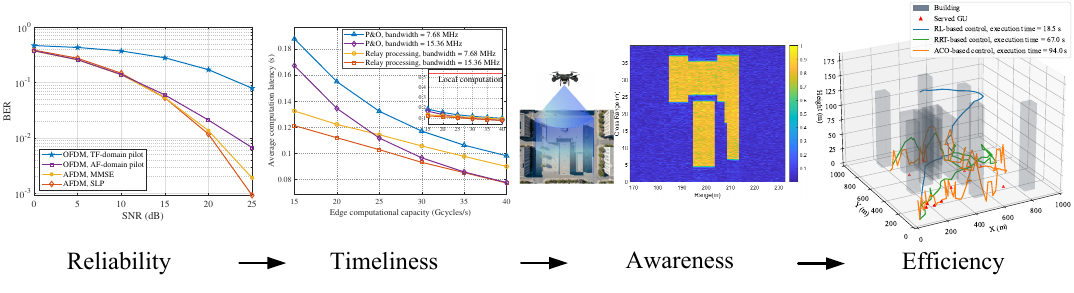}
	\caption{Case study results of the $\text{SC}^3$ closed-loop LAEI networks.}
	\label{simulation}
	\vspace{-0.5cm}
\end{figure*}

\section{Case Study}
To validate the practical efficacy of the proposed $\text{SC}^3$ based embodied intelligence framework, we consider a low-altitude mission in an unknown urban area of $1000~\mathrm{m}\times1000~\mathrm{m}$. In this scenario, a UAV continuously provides communication services for 10 GUs while navigating through a complex, obstacle-ridden area, where each GU requests a payload of $10$ Mbits. This closed-loop setting couples the requirements of reliable communication, timely computation, situational sensing awareness, and efficient control for safe and responsive operation. The UAV operates at a carrier frequency of 5.8 GHz with a subcarrier spacing of 120 kHz, flying at speeds up to 40 m/s \cite{luo2025trajectory}. In particular, the UAV reuses the OFDM based communication signals transmitted to GUs for SAR based environmental imaging \cite{10715676}. To support heterogeneous sensing and communication requirements, a frequency-division multiplexing scheme is adopted, where a wideband signal synthesized from 1200 subcarriers is allocated for high-resolution imaging and high-throughput communication for downlink users, while a smaller set of 128 subcarriers is assigned to UAV-edge server communication, as the locally pre-processed sensing data can generally be effectively compressed.

The results corresponding to the four parts of the $\text{SC}^3$ framework are illustrated in Fig.~\ref{simulation}, demonstrating how reliable transmission and efficient computation translate into high-fidelity sensing and intelligent reaction.
\begin{itemize}
	\item \textbf{Reliable Communication as the Foundation}: The first plot in Fig.~\ref{simulation} compares the bit error rate (BER) of the two transmission schemes proposed in Section~\ref{teccomm} with conventional OFDM and AFDM schemes under high-mobility conditions. Conventional OFDM with TF-domain pilots performs poorly due to severe Doppler-induced inter-carrier interference at 40 m/s. In contrast, AF-domain pilot-assisted OFDM substantially enhances link reliability, achieving BER performance comparable to that of AFDM with minimum mean square error (MMSE) equalization at SNRs below 15~dB, with significantly lower symbol detection complexity and flexible resource scheduling. As for downlink control signaling delivering, AFDM with symbol-level precoding (AFDM-SLP) scheme \cite{shuntian2026doppler}, through leveraging the full diversity of the AF-domain channel and turning the ICI into constructive components at the transmitter, improves reliability while lowering the demodulation effort required at the UAV receiver. 
	\item \textbf{Computational Coordination for Real-Time Response}: With the communication link established, the second plot in Fig.~\ref{simulation} illustrates how average task latency varies with computational capacity for the two orchestration strategies—packaging and offloading (P\text{\&}O) versus relay processing—under different transmission bandwidths. Here, the reported latency is the closed-loop response latency, accounting for sensing data transmission and processing as well as control-strategy generation and delivering. Compared with local computation, which yields an average latency of 0.541~s, UAV--edge coordination consistently keeps the closed-loop response latency below 0.20~s. The most significant improvement is observed at high edge computation capacity with sufficient bandwidth, where the latency decreases to 0.077~s at 40~Gcycles/s with 15.36~MHz for both strategies, corresponding to an approximately 85.8\% reduction relative to local computation. At lower edge computational capacities, relay processing exhibits lower latency, as it progressively forwards intermediate results. The gap is particularly pronounced when the transmission bandwidth is limited. As edge computational resources and transmission bandwidth increase, the latency gap narrows or even vanishes, and the P\text{\&}O scheme becomes more efficient, as it does not require the UAV, edge, and cloud to align their processing modules.
	\item \textbf{High-Fidelity Imaging via Collaborative Processing:} Based on the reliable transmission and coordinated computation strategies, the third plot presents the environment reconstruction results. By synthesizing a virtual aperture along its trajectory, the UAV successfully reconstructs fine-grained building contours and spatial layouts, achieving range and azimuth resolutions of 1~m and 0.5~m, respectively, under the parameter configuration specified earlier. The normalized SAR energy intensity closely matches the actual building shapes, which verifies that OFDM-based SAR can provide accurate and structured environmental representations. Such high-fidelity reconstructions further support closed-loop control tasks, e.g., the trajectory design illustrated in the last plot of Fig.~\ref{simulation}.
	\item \textbf{Intelligent Control through RL:} The last plot in Fig.~\ref{simulation} compares the three-dimensional UAV trajectories generated by three control schemes. Leveraging the integrated $\text{SC}^3$ feedback, the RL-based control achieves a superior mission completion time of 18.5~s. Unlike the heuristic competitors, the RL agent transforms high-fidelity sensing and reliable communication into actionable intelligence, rapidly ascending to high altitudes to bypass dense obstacles, hence resulting in a smooth, goal-oriented path that eliminates redundant detours. In contrast, rapidly-exploring random tree (RRT) \cite{wang2021trajectory} and ant colony optimization (ACO) rely solely on local, incremental, or iterative path search, resulting in conservative low-altitude navigation and multiple detours that significantly extend mission duration to 67 s and 94 s, respectively.
\end{itemize}

\section{Challenges and Future Research Directions}

As the integration of $\text{SC}^3$ forms the cornerstone for realizing LAEI, significant challenges remain in its early-stage development. Building on the foregoing discussion, major obstacles remain in high-fidelity sensing, high-speed communication, efficient computation, and precision control. This section outlines the key challenges and potential research directions in these critical areas.

\subsection{High-Fidelity Multi-Modal Sensing}

High-resolution sensing, especially through SAR imaging, is indispensable for providing the fine-grained spatial detail required for autonomous navigation. However, the pursuit of decimeter-level imaging resolution imposes prohibitive bandwidth and power demands, straining the onboard resources of the UAV. Furthermore, other single-modality sensing techniques are often insufficient for robust environmental mapping under diverse weather or lighting conditions. Therefore, a critical research direction lies in multi-modal perceptual synergy. The challenge therein is to develop sophisticated fusion frameworks that can harmonize heterogeneous data—such as optical, radar, and acoustic signals—to achieve concurrent high-precision localization and target recognition. Future research should explore how to prioritize task-relevant information within these multi-modal streams, enabling the system to maintain high perceptual fidelity, while minimizing the computational and spectral overhead of raw data processing.

\subsection{Semantic-Aware High-Speed Communication}

The ``perception-to-control" loop in LAEI networks generates a huge amount of sensing data, creating a fundamental conflict with the inherent power and spectral limitations of low-altitude platforms. Traditional bit-level communications are increasingly ill-suited for these high-throughput, low-latency requirements, leading to backhaul congestion and degraded responsiveness. Semantic communication (SemCom) offers a transformative solution. By leveraging task-oriented source compression, SemCom can extract and transmit only essential features, such as target contours or motion vectors, significantly reducing redundancy. 

Furthermore, SemCom is inherently synergistic with multi-modal sensing. Since both paradigms rely on the extraction and merging of high-level features—such as aligning radar point clouds with visual descriptors—SemCom can leverage the results of multi-modal fusion to create a highly compressed, consolidated representation of the environment. Nonetheless, several challenges remain, including defining unified task-specific semantics across different modalities and dynamically adjusting compression strategies in response to time-varying network conditions, and coordinating with sensing and control objectives to maintain overall system performance. Future research should explore adaptive semantic encoding and intelligent coordination schemes between sensing and transmission to ensure the $\text{SC}^3$ loop remains resilient in large-scale LAEI deployments.

\subsection{Wide-Area Elastic Computation}

As UAV deployments scale and sensing capabilities evolve, LAEI networks face an exponential surge in computational demands. Processing massive perception data across expansive geographical areas, especially in high-density scenarios, presents a fundamental challenge in resource orchestration. The core difficulty lies in designing a wide-area computational scheduling mechanism that dynamically allocating elastic resources across the UAV-edge-cloud continuum to satisfy the diverse requirements of full-scenario missions. 

Future research must move towards a large-scale collaborative, multi-tier computation framework, intelligently partitioning sensing tasks based on their spatial-temporal urgency and instantaneous load of heterogeneous nodes. The objective is to achieve a scalable and flexible computational architecture that can accommodate an unprecedented density of tasks without localized congestion, maintaining millisecond-level responsiveness.

\subsection{Precision Control and Adaptive Reasoning}

In complex and stochastic environments of LAEI networks, generating reliable control strategies remains a significant challenge. Current approaches built upon lightweight models often suffer from poor cross-scenario generalization, only performing well in environments that strictly mirror their training data. This limitation is exacerbated when UAVs operate across diverse mission contexts with varying obstacles, user distributions, and environmental conditions. Furthermore, lightweight models may fail to capture intricate dynamics needed for long-horizon decision-making, leading to suboptimal trajectory planning or task execution.

To address these challenges, two promising research frontiers emerge:
\begin{itemize}
\item \textbf{Transfer Learning for Multi-Scenario Adaptability:} By leveraging domain adaptation and knowledge transfer, models trained in well-defined environments can be swiftly adapted for unseen contexts. This enables UAVs to maintain high control precision with minimal retraining.
\item \textbf{Large Model-Driven Reasoning:} Beyond simple reactive control, the integration of Large Foundation Models and ``Large-Small" model coordination architecture offers a transformative path for complex inference. These models can provide superior reasoning and zero-shot prediction capabilities, allowing UAVs to make context-aware decisions in highly heterogeneous scenarios.
\end{itemize}
Developing adaptive decision-making strategies that combine model scalability with real-time responsiveness is essential for fulfilling the promise of embodied intelligence, ensuring that LAEI platforms remain resilient and precise in ever-changing operational environments.

\section{Conclusion}
This article presented an $\text{SC}^3$-based holistic framework for LAEI, elucidating the endogenous synergy between sensing, communication, computation, and control. We overviewed the critical techniques in these four aspects, and the case study demonstrated that by integrating Doppler-resilient transmission and hierarchical computation orchestration, an RL-driven embodied agent can efficiently translate high-fidelity SAR imaging into goal-oriented trajectories. These insights emphasize both the promise and the challenges of LAEI, providing a strategic roadmap for overcoming the resource constraints of low-altitude environments.

\bibliographystyle{IEEEtran}%
\bibliography{bibfile.bib}

\vfill

\end{document}